\documentclass[nofootinbib,aps,pre,twocolumn,superscriptaddress,showkeys,showpacs,longbibliography]{revtex4-1}
\usepackage{graphicx}
\usepackage{color}
\usepackage{soul}

\usepackage{amsmath,amssymb}

\begin{document}

\title{Phototactic Robot Tunable by Sensorial Delays}

\author{Maximilian Leyman}
\thanks{These two authors contributed equally.}
\affiliation{Department of Physics, University of Gothenburg, SE-41296 Gothenburg, Sweden, EU}
\affiliation{Department of Physics, Chalmers University of Technology, SE-41296 Gothenburg, Sweden, EU}

\author{Freddie Ogemark}
\thanks{These two authors contributed equally.}
\affiliation{Department of Physics, University of Gothenburg, SE-41296 Gothenburg, Sweden, EU}
\affiliation{Department of Physics, Chalmers University of Technology, SE-41296 Gothenburg, Sweden, EU}

\author{Jan Wehr}
\affiliation{Department of Mathematics and Program in Applied Mathematics, University of Arizona, Tucson, Arizona 85721, USA}

\author{Giovanni Volpe}
\affiliation{Department of Physics, University of Gothenburg, SE-41296 Gothenburg, Sweden, EU}

\date{\today}

\begin{abstract}
The presence of a delay between sensing and reacting to a signal can determine the long-term behavior of autonomous agents whose motion is intrinsically noisy.
In a previous work [M. Mijalkov, A. McDaniel, J. Wehr, and G. Volpe, Phys. Rev. X 6, 011008 (2016)], we have shown that sensorial delay can alter the drift and the position probability distribution of an autonomous agent whose speed depends on the illumination intensity it measures. 
Here, using theory, simulations, and experiments with a phototactic robot, we generalize this effect to an agent for which both speed and rotational diffusion depend on the illumination intensity and are subject to two independent sensorial delays. 
We show that both the drift and the probability distribution are influenced by the presence of these sensorial delays.  In particular, the radial drift may have positive as well as negative sign, and the position probability distribution peaks in different regions depending on the delay.
Furthermore, the presence of multiple sensorial delays permits us to explore the role of the interaction between them.
\end{abstract}


\maketitle

\section{Introduction}\label{sec:intro}

Autonomous robots are increasingly being employed both in fundamental research and in technological settings \cite{schweitzer2003brownian}. 
One of the critical tasks in their development is to make them capable of complex autonomous behaviors in response to environmental cues, while keeping their hardware, sensorial inputs and software as simple as possible \cite{rubenstein2014programmable,werfel2014designing}.
In fact, complex behaviors emerging from agents obeying simple rules have the advantage of being extremely robust and reliable \cite{schweitzer2003brownian,palacci2013living,chepizhko2013diffusion}.
Often, a source of inspiration are the behaviors of simple organisms like foraging insects \cite{bonabeau2000inspiration} and chemotactic bacteria \cite{berg2008coli}.

Usually, the robots are designed to react to real-time sensorial inputs from their surroundings and make decisions based on this information. 
In Nature, however, there are several examples of microscopic organisms and animals that compare current information about their surroundings with previous information, and adjust their behavior by making extrapolations. 
For example, chemotactic bacteria have been shown to adjust their motion by comparing the chemical concentration in their surroundings at different times \cite{macnab1972gradient,segall1986temporal}, and  
insects, fishes and humans extrapolate their positions forward in time when navigating in groups \cite{collett1978hoverflies,nijhawan1994motion,rossel2002predicting}. 
These behaviors result in the introduction of a sensorial delay between the sensorial input perception and the ensuing behavioral response.
We have recently explored the role played by this sensorial delay both theoretically and experimentally \cite{mijalkov2016engineering,volpe2016effective}. Using a phototactic robot whose speed depended on the  measured light intensity, we demonstrated that introducing a sensorial delay could make the robot either stay near or avoid the light source; furthermore, when multiple light-emitting robots interacted, we showed that this effect promoted either aggregation or segregation. 
The presence of negative sensorial delay, sometimes called ``anticipation", has also been shown to greatly influence the dynamics of a system of interacting agents and the patterns that are formed \cite{gerlee2017impact}, as well as to affect the clustering tendencies of agents in a two-dimensional variant of the Vicsek model \cite{piwowarczyk2018influence}. 

Here, using theory, simulations, and experiments with phototactic robots, we generalize the effect we described in Ref.~\cite{mijalkov2016engineering} to the case of an agent whose speed and rotational diffusion depend on the illumination intensity and are subject to two independent sensorial delays. 
Using a phototactic robot moving within an arena illuminated with a radial light intensity pattern, we investigate how the robot's behavior is affected by a delay when only its speed varies as a function of the intensity, only its rotational diffusion varies, or both quantities vary simultaneously. 
We show that both its drift and its position probability distribution are influenced by the presence of these sensorial delays.  In particular, the radial drift may have positive as well as negative sign, and the position probability distribution may peak in different regions depending on the delay.
The presence of multiple sensorial delays permits us to explore the role of the interaction between them.

\section{Model}\label{sec:model}

\begin{figure}
\includegraphics[width=1.0\columnwidth]{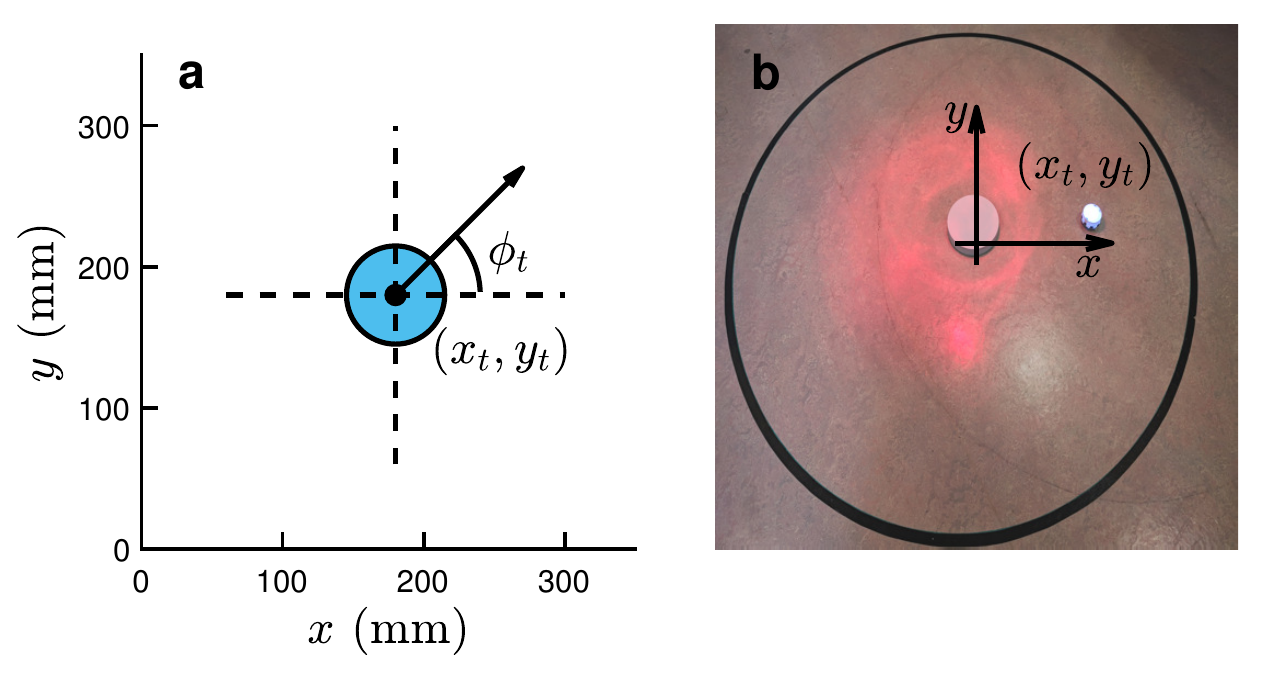}
\caption{
Model and experimental setup.
(a) The robot is at position $(x_t,y_t)$ at time $t$ and moves at a speed $v$ with orientation $\phi_t$ as indicated by the arrow. This orientation is subject to a noise so that the robot's characteristic reorentiation time is $\tau$.
(b) Picture of the robot in the arena illuminated with a light field generated by an infrared lamp. The robot is free to move in the region between the round object placed at the center of the arena and the black line on the outer edge of the arena. Depending on the scenario, the robot will either modify its speed (Fig.~\ref{fig2}), its rotational diffusion (Fig.~\ref{fig3}), or both simultaneously (Figs.~\ref{fig4} and \ref{fig5}) as functions of the light intensity it measures.
}
\label{fig1}
\end{figure}

The robot we employ can be modelled as an autonomous agent performing active Brownian motion \cite{volpe2014simulation}: it moves in the $xy$-plane while its orientation is subject to noise (Fig.~\ref{fig1}a). Its behavior can therefore be modeled by the following stochastic differential equations (SDEs) \cite{oksendal2003stochastic}:
\begin{equation}\label{eq:ABM}
    \begin{gathered}
    \begin{cases}
        \displaystyle \frac{dx_t}{dt} = v\cos{\phi_{t}}\\[10pt]
        \displaystyle \frac{dy_t}{dt} = v\sin{\phi_{t}}\\[10pt]
        \displaystyle \frac{d\phi_t}{dt} = \sqrt{2 \over \tau}~\eta_t
        \end{cases}
    \end{gathered}
\end{equation}
where $(x_t,y_t)$ is the robot's position at time $t$, $\phi_t$ is its orientation, $v$ is its speed, $\tau$ is its characteristic reorientation time (i.e., the time during which its orientation varies on average by one radiant), and $\eta_t$ is a normally distributed white noise term with zero mean and unit intensity. 

Let us assume that the arena where the robot moves is illuminated by a light intensity $I(x,y)$. If the robot can measure $I$ and react to this measurement by adjusting its speed $v$ and rotational diffusion $R$, the SDEs describing its motion become:
\begin{equation}\label{eq:ABM:I}
    \begin{gathered}
    \begin{cases}
        \displaystyle \frac{dx_t}{dt} = v(I_t)\cos{\phi_{t}}\\[10pt]
        \displaystyle \frac{dy_t}{dt} = v(I_t)\sin{\phi_{t}}\\[10pt]
        \displaystyle \frac{d\phi_t}{dt} = \sqrt{2R(I_t) \over \tau}~\eta_t
        \end{cases}
    \end{gathered}
\end{equation}
where $I_t = I(x_t,y_t)$.
As in our previous work \cite{mijalkov2016engineering}, we will assume that the robot speed is bounded by a minimum speed $v_{\rm min}$ and a maximum speed $v_{\rm max}$, and decreases with higher light intensity, i.e.:
\begin{equation}\label{eq:speedDependency}
    v(I) = v_{\rm min} + (v_{\rm max}-v_{\rm min})e^{-I}.
\end{equation}
Furthermore, we now vary also the normalized rotational diffusion coefficient, so that it is bounded between minimum and maximum values $R_{\rm min}$ and $R_{\rm max}$, and increases with higher light intensities, i.e.:
\begin{equation}\label{eq:DiffusionDependent}
    R(I)=R_{\rm max}-(R_{\rm max}-R_{\rm min})e^{-I}.
\end{equation}

We finally introduce the sensorial delays so that SDEs~(\ref{eq:ABM:I}) become the stochastic differential delay equations (SDDEs):
\begin{equation}\label{eq:ABM:delay}
    \begin{gathered}
    \begin{cases}
        \displaystyle \frac{dx_t}{dt} = v(I_{t-\delta_{\rm v}})\cos{\phi_{t}}\\[10pt]
        \displaystyle \frac{dy_t}{dt} = v(I_{t-\delta_{\rm v}})\sin{\phi_{t}}\\[10pt]
        \displaystyle \frac{d\phi_t}{dt} = \sqrt{2R(I_{t-\delta_{\rm R}}) \over \tau}~\eta_t
        \end{cases}
    \end{gathered}
\end{equation}
where $\delta_{\rm v}$ is the sensorial delay of the adjustment of the speed and $\delta_{\rm R}$ is the sensorial delay of the adjustment of the rotational diffusion coefficient.
A positive delay value corresponds to a delay in the time it takes to react to sensorial input, while a negative value corresponds to making a prediction of a future measured intensity \cite{mijalkov2016engineering,volpe2016effective,gerlee2017impact}.

\section{Theory}\label{sec:theory}

We theoretically study SDDEs~(\ref{eq:ABM:delay}) using multiscale analysis and derive expressions for the drift and steady-state position probability distribution of the robot. 
The multiscale analysis is a homogenization technique that is performed by taking to zero the characteristic time scales of the processes involved in determining the dynamics of the system, while keeping their ratios constant \cite{pavliotis2008multiscale}.
The detailed derivations are provided in Appendix~\ref{appendix}, while here we provide only an outline of the derivation and the key theoretical results.

\subsection{Outline of the derivation}

We start by rewriting SDDEs~(\ref{eq:ABM:delay}) in a more convenient form for the theoretical analysis, introducing, in particular, a parameter $\epsilon$ that will be taken to zero in the multiscale analysis \cite{pavliotis2008multiscale}.
We note that the speed of the robot is a function of its position, i.e., $v(x,y)=v(I(x,y))$, and that the robot changes the direction of its velocity according to a random process, at a rate which is also a function of the position, i.e., $\sigma(x,y) = \sqrt{2R(I(x,y))/\tau}$.
If the robot reacts to the environment with a delay $\delta_{\rm v} = c\epsilon^2$, the speed at time $t$ is proportional to $v\left(x_{t-c\epsilon^2}, y_{t-c\epsilon^2}\right)$ (the value of the function $v$ evaluated at the position of the particle at an earlier moment of time, if $c>0$, or at a later moment, if $c<0$).  
Likewise, the rate of the robot random rotation is proportional to $\sigma\left(x_{t-k\epsilon^2}, y_{t-k\epsilon^2}\right)$, with a delay $\delta_{\rm R} = k\epsilon^2$.
The parameters $c$ and $k$ are constants, positive or negative, and, in general, different from one another.
Thus, we can rewrite SDDEs~(\ref{eq:ABM:delay}) as a set of SDDEs with a small parameter $\epsilon$:
\begin{equation}\label{eq:SDDE}
    \begin{gathered}
    \begin{cases}
        \displaystyle dx_t = {1 \over \epsilon}v\left(x_{t-c\epsilon^2}, y_{t-c\epsilon^2}\right)\cos \phi_t~dt\\[10pt]
        \displaystyle dy_t = {1 \over \epsilon}v\left(x_{t-c\epsilon^2}, y_{t-c\epsilon^2}\right)\sin \phi_t~dt\\[10pt]
        \displaystyle d\phi_t = {1 \over \epsilon}\sigma\left(x_{t-k\epsilon^2}, y_{t-k\epsilon^2}\right)~dW_t
        \end{cases}
    \end{gathered}
\end{equation}
where $W_t$, $t \geq 0$, denotes a Wiener process.\footnote{The stochastic differential in the third equation can also be written as $dW_t = \eta_t\,dt$, where $\eta_t$, $t \geq 0$, is a unit white noise process.}
Since the factor of ${1 \over \epsilon}$ in the equation for $\phi_t$ makes the changes of direction occur very rapidly for $\epsilon$ small, we scale the speed $v$ in the first two equations in the same way to obtain a nontrivial limiting dynamics for the position of the robot.
We remark that the SDDEs~(\ref{eq:SDDE}) becomes SDDEs~(\ref{eq:ABM:delay}) for $\epsilon=1$.

We study the limit of SDDEs~(\ref{eq:SDDE}) for $\epsilon \to 0$, which is equivalent to accelerating the microscopic dynamics (speed, rotation, delays) of the system while keeping its macroscopic properties (drift, probability distribution) fixed.
We first linearize $x$ and $y$ as functions of time, and then $v$ and $\sigma$ as functions of $x$ and $y$, to approximate the SDDEs~(\ref{eq:SDDE}) by a system of SDEs without delays.
We then consider the corresponding backward Kolmogorov equation for the probability density $\rho$,
write the function $\rho$ as a formal series in powers of $\epsilon$, i.e.,
\begin{equation}\label{eq:rho}
\rho = \rho_0 + \epsilon \rho_1 + \epsilon^2 \rho_2 + \dots,
\end{equation}
and use the multiscale expansion method to derive the backward Kolgomorov equation for the limiting density $\rho_0$:
\begin{align}\label{eq:rho0}
{\partial \rho_0 \over \partial t} 
= 
&-{1 \over 2} 
{\delta_v \over \tau}{2R \over \sigma^2} %
v 
\left\{ v_x {\partial \rho_0 \over\partial x} + v_y {\partial \rho_0 \over \partial y} \right\} \cr
&+ 
{\delta_R \over \tau}{2R \over \sigma^2} %
{v^2 \over \sigma} 
\left\{ \sigma_x {\partial \rho_0 \over \partial x} +  \sigma_y {\partial \rho_0 \over \partial y} \right\} \cr
&+ v 
\left\{
{\partial\over\partial x}\left[{v \over \sigma^2}{\partial\over\partial x}\rho_0\right] + {\partial\over\partial y}\left[{v \over \sigma^2}{\partial\over\partial y}\rho_0\right]
\right\}.
\end{align}
From this equation, we derive the limiting SDEs satisfied by the processes $x_t$ and $y_t$:
\begin{equation}\label{eq:xy}
    \begin{gathered}
    \begin{cases}
        \displaystyle dx_t = 
        & \displaystyle \left[-{1 \over 2} 
        {\delta_v \over \tau}{2R \over \sigma^2}  %
        vv_x + 
	{\delta_R \over \tau}{2R \over \sigma^2} %
        {v^2 \over \sigma}\sigma_x + v{\partial\over\partial x}\left({v \over \sigma^2}\right)\right]\,dt \\[10pt]
        & \displaystyle + \sqrt{2}{v \over \sigma}\,dW_t^{(1)} \\[10pt]
        \displaystyle dy_t = 
        & \displaystyle \left[-{1 \over 2} 
        {\delta_v \over \tau}{2R \over \sigma^2} %
        vv_y + 
        {\delta_R \over \tau}{2R \over \sigma^2} %
        {v^2 \over \sigma}\sigma_y + v{\partial\over\partial y}\left({v \over \sigma^2}\right)\right]\,dt \\[10pt]
        & \displaystyle + \sqrt{2}{v \over \sigma}\,dW_t^{(2)}
        \end{cases}
    \end{gathered}
\end{equation}
where $W^{(1)}$ and $W^{(2)}$ are independent Wiener processes.

From SDEs~(\ref{eq:xy}), we obtain the associated forward Kolmogorov (or  Fokker-Planck) equation
\begin{equation}\label{eq:FP}
\partial_t \rho_0 = 
{1 \over 2} 
{\delta_v \over \tau}{2R \over \sigma^2} %
\nabla\cdot\left(\rho_0 v\nabla v\right) 
- 
{\delta_R \over \tau}{2R \over \sigma^2} %
\nabla\cdot\left(\rho_0{v^2 \over \sigma}\nabla\sigma\right) 
+ \nabla\cdot\left({v \over \sigma^2}\nabla\left(v\rho_0\right)\right),
\end{equation}
from which the stationary probability density $\rho_0$ can be found by solving for $\partial_t \rho_0 = 0$.

\subsection{Key results in circular geometry}

Given the circular geometry of our experiment (see Section~\ref{sec:robot}), we can assume $v$ and $\sigma$ in SDEs~(\ref{eq:xy}) and in Eq.~(\ref{eq:FP}) to be rotationally invariant. We can therefore study these equations in polar coordinates focusing specifically on the radial coordinate $r = \sqrt{x^2+y^2}$.\footnote{The results for the azimuthal coordinate are trivial: because of rotation symmetry, the azimuthal drift must be null and the azimuthal position probability distribution must be uniform.} 
We therefore obtain the following homogenized SDE for the radial coordinate:
\begin{equation}\label{eq:dr}
dr_t = d(r)\,dt + s(r)\,d\tilde{W}_t,
\end{equation}
where $\tilde{W}_t$ is a Wiener process, the radial drift coefficient is
\begin{align}\label{eq:D(r)}
d(r) = -{1 \over 2} 
{\delta_v \over \tau}{2R \over \sigma^2} %
vv_r + 
{\delta_R \over \tau}{2R \over \sigma^2} %
{v^2\sigma_r \over \sigma} + v\left({v \over \sigma^2}\right)_r + {1 \over r}{v^2 \over \sigma^2}
\end{align}
and the noise coefficient is
\begin{equation}
s(r) = \sqrt{2}{v(r) \over \sigma(r)}.
\end{equation}
The steady-state radial probability distribution is
\begin{equation}\label{eq:rho(r)}
\rho_0(r) = {B \over s^2(r)}\exp\left\{ \int {2b(r) \over s^2(r)}dr \right\},
\end{equation}
where $B$ has to be adjusted to make the integral of $\rho$ equal $1$.  

\section{Robot Experiment}\label{sec:robot}

The experimental setup is shown in Fig.~\ref{fig1}b.
We use an Elisa-3 \cite{Elisa-3} robot, which is an autonomous robot with a circular shape that measure $50~{\rm mm}$ in diameter and $30~{\rm mm}$ in height. 
The robot moves at a maximum speed of $60~{\rm cm\,s^{-1}}$ thanks to two wheels on either side powered by direct-current (DC) motors.
It is equipped with eight infrared (IR) sensors that measure ambient light placed along the perimeter of the robot at equal intervals of 45 degrees to create a detection field of 360 degrees.
Furthermore, the robot features proximity sensors that permit it to detect the presence of objects at a distance of $6~{\rm cm}$ and four ground sensors that permit it to detect the presence of a black border on the ground.

We have programmed the robot using Aseba studio \cite{Aseba}.
The robot can perform Brownian motion through a cycle of two phases: a ``forward phase" when the robot moves forward at constant speed along a straight line for $0.1~{\rm s}$; and a ``rotation phase" when the robot changes its direction by a random angle for $0.1~{\rm s}$.
This cycle is repeated to emulate a Brownian motion. 

We delimit a region where the robot can move freely by placing a circular object at the center of the arena and a black tape along its outer edge, as shown in Fig.~\ref{fig1}b.
The robot uses the proximity sensors to detect the circular object and the ground sensors to detect the black tape, and it avoids them by changing its direction away from them, i.e. until it does not detect their presence any more. 

We generate a radially decaying light intensity field by placing a 150-W IR lamp above the arena where the robot moves.
The robot measures the local value of this light intensity using the IR sensors and adapts its behavior accordingly.
Following the approach in our previous work \cite{mijalkov2016engineering}, we estimate the values of $I_{t-\delta_{\rm v}}$ and $I_{t-\delta_{\rm R}}$ by an expansion to the first order, i.e., $I(t-\delta_{\rm v})=I(t)-\delta_{\rm v} I'(t)$ and $I(t-\delta_{\rm R})=I(t)-\delta_{\rm R} I'(t)$, respectively. 
Practically, the robot stores the value of the intensity in the previous and current motion cycles, and uses them to approximate the intensity derivative.

During the experiments, the robot's positions are recorded with a videocamera at $32~{\rm fps}$ and tracked using standard digital video microscopy algorithms.
Each experiment runs for 60 minutes. 

From the acquired trajectories, we estimate the radial probability distribution of the robot's position and its radial drift. 
The radial probability distribution is the probability of finding a robot at a certain radial distance from the center of the arena and is directly measured from the histogram of the robot's positions.
The radial drift shows how the robot moves on average relative to the center of the arena depending on its location and is measured using the following equation \cite{mijalkov2016engineering}:
\begin{equation}
    d(r)=\frac{1}{\Delta t}\langle r_{n+1}-r_n | r_n\cong r\rangle,
\end{equation}
where $r_n$ is the series of robot's positions and $\Delta t$ is the time step.
If the radial drift is positive, the robot on average moves away from the center of the arena, whereas a negative drift means that it moves on average towards the center.

\section{Results}

We consider three scenarios. 
First, we vary only the speed as a function of light intensity (as in our previous work \cite{mijalkov2016engineering}).
Second, we vary the rotational diffusion coefficient.
Third, we vary both quantities simultaneously so that the presence of multiple sensorial delays permits us to explore how they interact.
In all cases, we present the theoretical, simulation and experimental results. 
The simulations are realized by a finite-difference algorithm \cite{volpe2014simulation} that implements SDDEs~(\ref{eq:ABM:delay}) using the experimental parameters.\footnote{As in the experiments, we introduce the sensorial delays by estimating the values of $I_{t-\delta_{\rm v}}$ and $I_{t-\delta_{\rm R}}$ by an expansion to the first order, i.e., $I(t-\delta_{\rm v})=I(t)-\delta_{\rm v} I'(t)$ and $I(t-\delta_{\rm R})=I(t)-\delta_{\rm R} I'(t)$, respectively.}

\subsection{Speed dependent on the light intensity}

\begin{figure*}
\includegraphics[width=1.0\textwidth]{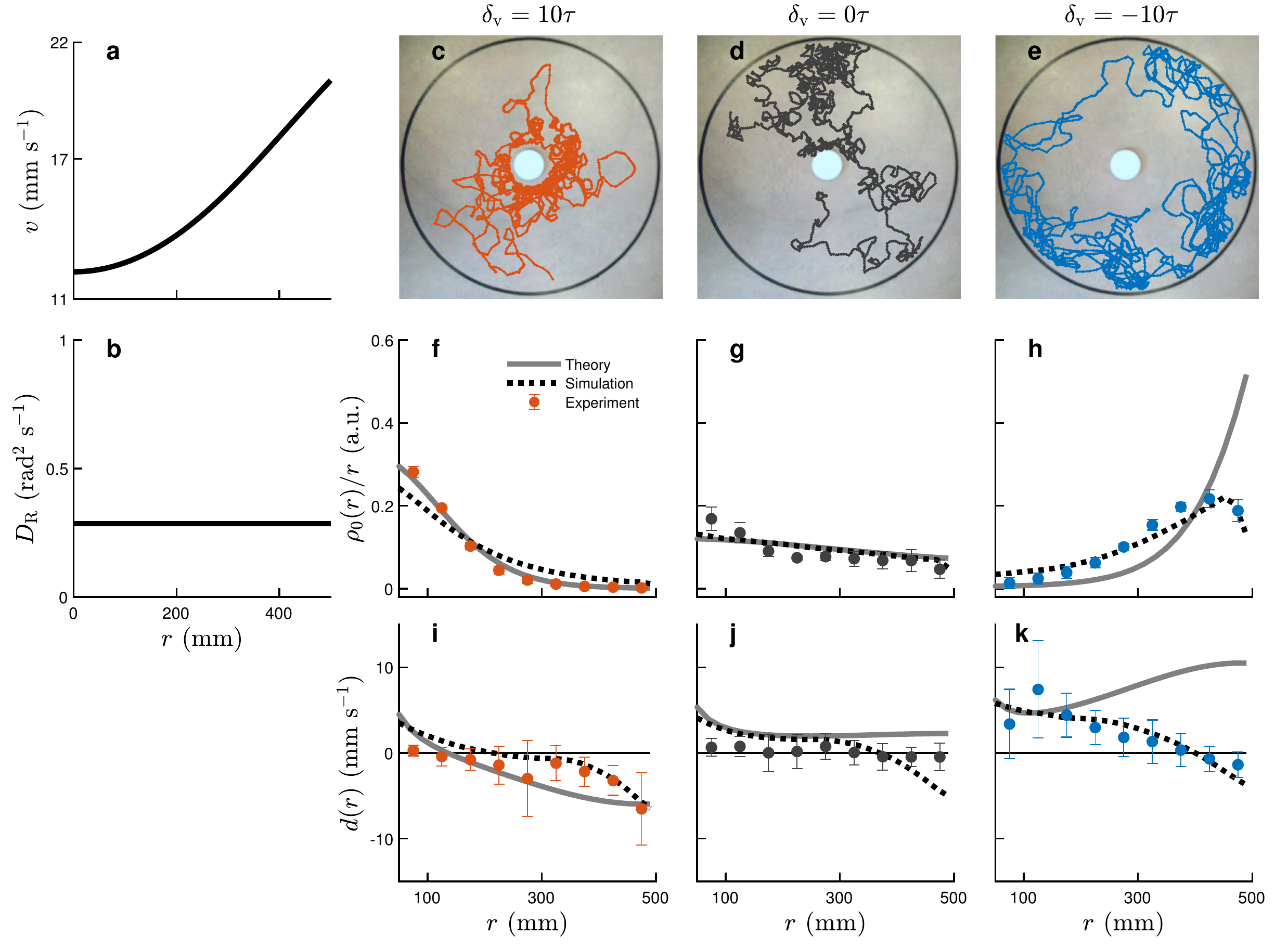}
\caption{
Robot behavior with sensorial delay in the speed.
(a) Speed $v(r)$ and (b) rotational diffusion coefficient $D_{\rm R}(r)$ as a function of radial position. 
(c-e) 60-minute-long trajectories of the robot within the arena for positive, zero and negative delays. 
(f-h) Radial probability distributions $\rho_0(r)/r$ and (i-k) radial drift $d(r)$ of the robot for positive, zero and negative delays; the symbols represent experimental data with standard deviation, the dashed lines represent simulations, and the solid lines represent the theory (Eqs.~(\ref{eq:D(r)}) and (\ref{eq:rho(r)})).
}
\label{fig2}
\end{figure*}

We set the speed to vary between the $v_{\rm max} = 25.7~{\rm mm\,s^{-1}}$ and $v_{\rm min} = 4.3~{\rm mm\,s^{-1}}$ according to Eq.~(\ref{eq:speedDependency}) (Fig.~\ref{fig2}a), while the rotational diffusion is kept constant at $D_{\rm R} = \tau^{-1} = 0.29~{\rm rad^2s^{-1}}$ (Fig.~\ref{fig2}b).
This case is equivalent to that we had previously studied \cite{mijalkov2016engineering}.

The qualitative behavior of the robot can be seen from its trajectories in the presence of different sensorial delays.
In the absence of any delay (Fig.~\ref{fig2}d), the robot has a slight preference to spend time in the regions with low speed (corresponding to high light intensity).
This tendency is accentuated when a positive sensorial delay is introduced ($\delta_{\rm v} = +10\tau$, Fig.~\ref{fig2}c), while it can be reversed by introducing a sufficiently large negative delay ($\delta_{\rm v} = -10\tau$, Fig.~\ref{fig2}e).

These qualitative observations can be made more precise by measuring the radial probability distribution $\rho_0(r)$ (Figs.~\ref{fig2}f-h) and the radial drift $d(r)$ (Figs.~\ref{fig2}i-k) of the robot in each case.
The theoretical results (solid lines) agree well with the simulations (dashed lines) and experiments (symbols).
As we qualitatively discussed above (Figs.~\ref{fig2}c-e), the sensorial delay $\delta_{\rm v}$ influences the robot probability distribution $\rho_0(r)$ so that $\rho_0(r)$ peaks in the regions with higher light intensity and lower speed for $\delta_{\rm v}=+10\tau$ (Fig.~\ref{fig2}f), and in the regions with lower light intensity and higher speed for $\delta_{\rm v}=-10\tau$ (Fig.~\ref{fig2}h).
The radial drift $d(r)$ is also influenced by $\delta_{\rm v}$ so that $d(r)$ is mostly negative when $\delta_{\rm v}=+10\tau$ pulling the robot towards the center of the arena (Fig.~\ref{fig2}i), and it is positive when $\delta_{\rm v}=-10\tau$ pushing the robot towards the edge of the arena (Fig.~\ref{fig2}k).\footnote{Note that in this case (i.e. constant rotational diffusion) the critical value where the sign change of $d(r)$ occurs is $\delta_{\rm v} = -{2\over\sigma}$, as we have shown in Ref.~\cite{mijalkov2016engineering}.}

We observe that there are significant deviations between the theoretical $\rho_0(r)$ and $d(r)$ and those obtained from experiments and simulations, especially towards the edges of the arena. These deviations emerge because the experiments and simulations implement SDDEs~(\ref{eq:ABM:delay}), corresponding to $\epsilon = 1$ in SDDEs~(\ref{eq:SDDE}), while the theory is strictly valid for $\epsilon \to 0$. This is discussed in more detail in Section~\ref{sec:tvs}.

\subsection{Rotational diffusion dependent on the light intensity}

\begin{figure*}
\includegraphics[width=1.0\textwidth]{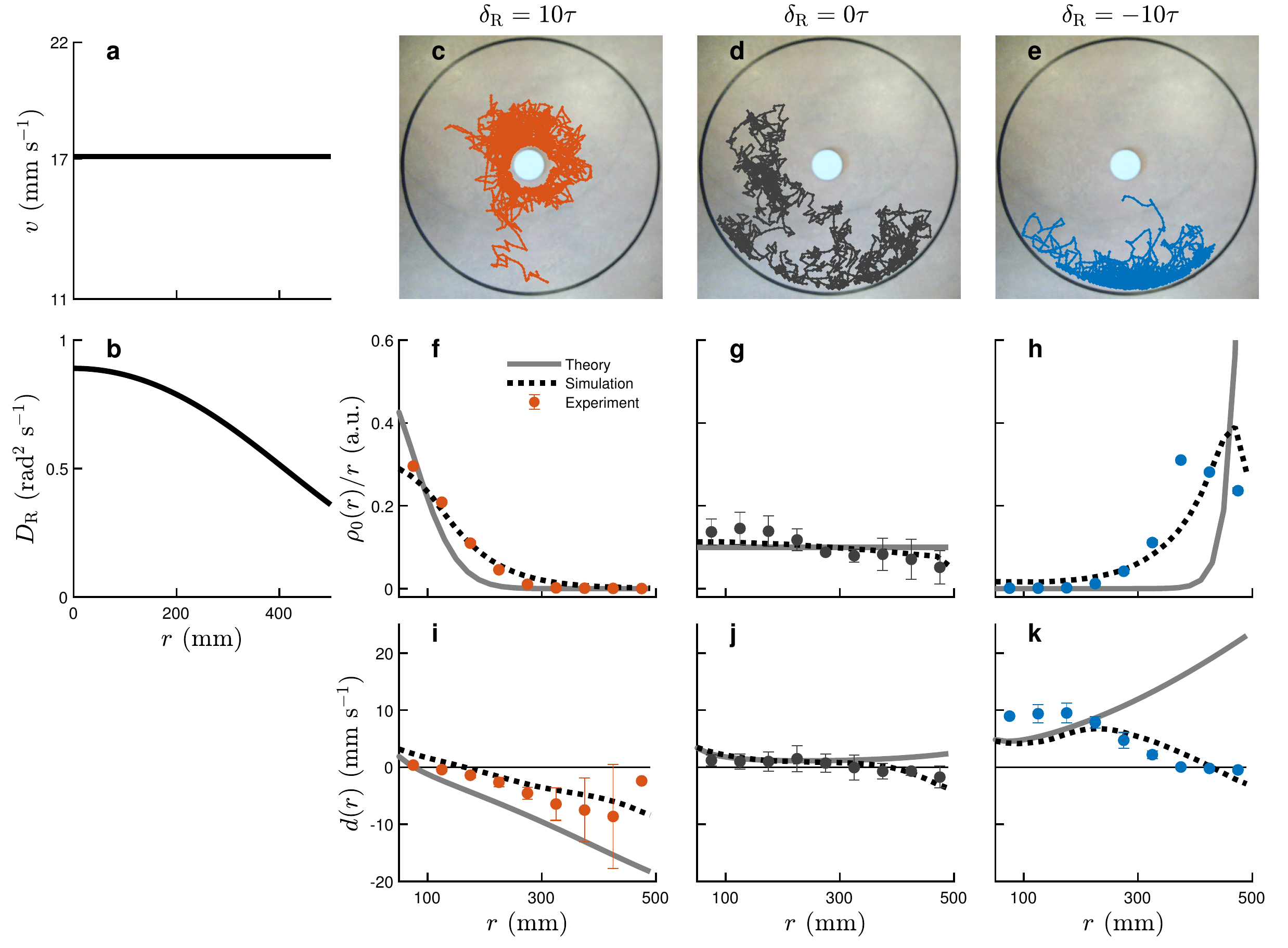}
\caption{
Robot behavior with sensorial delay in the rotational diffusion.
(a) Speed $v(r)$ and (b) rotational diffusion coefficient $D_{\rm R}(r)$ as a function of radial position. 
(c-e) 60-minute-long trajectories of the robot within the arena for positive, zero and negative delays. 
(f-h) Radial probability distributions $\rho_0(r)/r$ and (i-k) radial drift $d(r)$ of the robot for positive, zero and negative delays; the symbols represent experimental data with standard deviation, the dashed lines represent simulations, and the solid lines represent the theory (Eqs.~(\ref{eq:D(r)}) and (\ref{eq:rho(r)})).
}
\label{fig3}
\end{figure*}

In this second case, we set the rotational diffusion to vary between $D_{\rm R,max} =1.4~{\rm rad^2 s^{-1}}$ and $D_{\rm R,min} = 0.014~{\rm rad^2s^{-1}}$ (Fig.~\ref{fig3}b) according to Eq.~(\ref{eq:DiffusionDependent}), while keeping $v = 17.1~{\rm mm\,s^{-1}}$ (Fig.~\ref{fig3}a).

Figs.~\ref{fig3}c-e show the trajectories of the robot for various sensorial delays.
For positive delay ($\delta_{\rm R} = +10\tau$, Fig.~\ref{fig3}c), the robot spends most of its time close to the center of the arena, where the light intensity and the rotational diffusion take larger values.
For zero delay ($\delta_{\rm R} = 0$, Fig.~\ref{fig3}d), in full agreement with Eq.~(\ref{app37}), the space explored by the robot does not seem to be influenced by the light intensity and rotational diffusion values.
For negative delay ($\delta_{\rm R} = -10\tau$, Fig.~\ref{fig3}e), the robot spends most of its time in the region close to the edge of the arena, where the light intensity and the rotational diffusion take smaller values.

The radial probability distribution $\rho_0(r)$ (Figs.~\ref{fig3}f-h) and the radial drift $d(r)$ (Figs.~\ref{fig3}i-k) confirm these qualitative observations. In particular, we observe that the theoretical $\rho_0(r)$ when $\delta_{\rm R}=0$ corresponds to a uniform distribution (solid line in Fig.~\ref{fig3}g), and the corresponding $d(r)$ is almost zero (solid line in Fig.~\ref{fig3}j). 
For $\delta_{\rm R}>0$, $\rho_0(r)$ is peaked towards the high intensity and rotational diffusion regions near the center of the arena (Fig.~\ref{fig3}f), and $d(r)$ assumes mostly negative values, pulling the robot towards the arena center  (Fig.~\ref{fig3}i).
For $\delta_{\rm R}>0$, the reverse is true: $\rho_0(r)$ is peaked towards the low intensity and rotational diffusion regions near the edge of the arena (Fig.~\ref{fig3}f), and $d(r)$ assumes positive values, pushing the robot towards the arena edge (Fig.~\ref{fig3}i).\footnote{Note that in this case (i.e. constant speed) the critical value where the sign change of $d(r)$ occurs is $\delta_{\rm R} = 0$.}

Also in this case there are some discrepancies between the theory (gray lines) and the simulations (dashed lines) and experiments (symbols), which can be explained by the fact that simulations and experiments are not realized at the limit for $\epsilon \to 0$ (see Section~\ref{sec:tvs}).

\subsection{Both speed and rotational diffusion dependent on the light intensity}

\begin{figure*}
\includegraphics[width=1.0\textwidth]{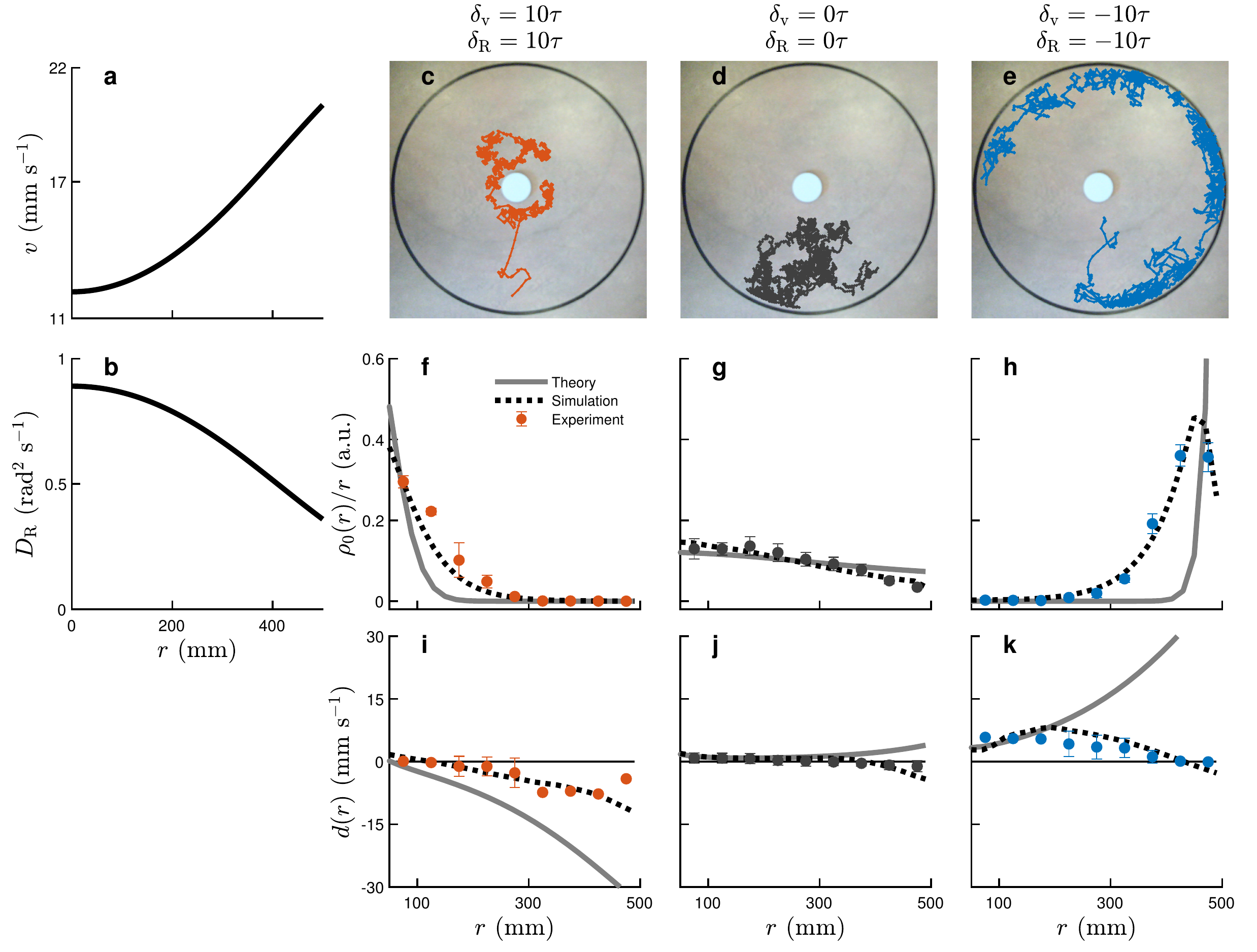}
\caption{
Robot behavior with equal sensorial delays in both the speed and the rotational diffusion.
(a) Speed $v(r)$ and (b) rotational diffusion coefficient $D_{\rm R}(r)$ as a function of radial position. 
(c-e) 60-minute-long trajectories of the robot within the arena for positive, zero and negative delays. 
(f-h) Radial probability distributions $\rho_0(r)/r$ and (i-k) radial drift $d(r)$ of the robot for positive, zero and negative delays; the symbols represent experimental data with standard deviation, the dashed lines represent simulations, and the solid lines represent the theory (Eqs.~(\ref{eq:D(r)}) and (\ref{eq:rho(r)})).
}
\label{fig4}
\end{figure*}

\begin{figure*}
\includegraphics[width=.75\textwidth]{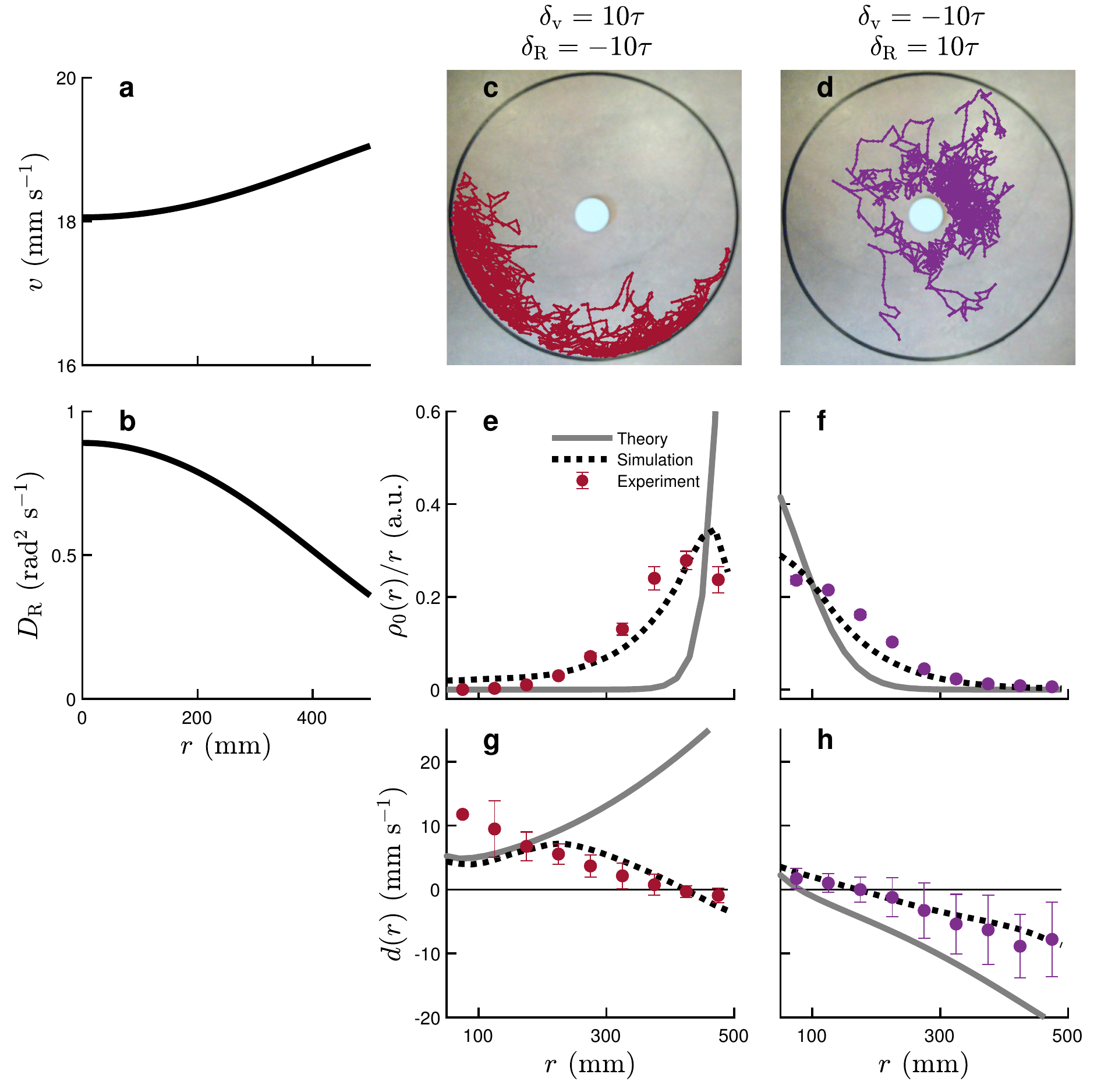}
\caption{
Robot behavior with opposite sensorial delays in the speed and the rotational diffusion.
(a) Speed $v(r)$ and (b) rotational diffusion coefficient $D_{\rm R}(r)$ as a function of radial position. 
(c-d) 60-minute-long trajectories of the robot within the arena. 
(e-f) Radial probability distributions $\rho_0(r)/r$ and (g-h) radial drift $d(r)$ of the robot; the symbols represent experimental data with standard deviation, the dashed lines represent simulations, and the solid lines represent the theory (Eqs.~(\ref{eq:D(r)}) and (\ref{eq:rho(r)})).
}
\label{fig5}
\end{figure*}

It is also interesting to consider the hybrid cases when both the speed and the rotational diffusion depend on the light intensity.
In Fig.~\ref{fig4}, we consider the case where both sensorial delays have the same sign, and reinforce each other. 
Figs.~\ref{fig4}a and \ref{fig4}b show $v(r)$ and $D_{\rm R}(r)$, respectively. 
Some samples of the resulting trajectories are shown in Figs.~\ref{fig4}c-e. 
The two sensorial delays reinforce each other and produce a more pronounced effect on the way the particle explores the space: 
when $\delta_{\rm v}=+10\tau$ and $\delta_{\rm R}=+10\tau$, the robot is attracted towards the high-light-intensity regions at the center of the arena where the speed is low and the rotational diffusion is high (Fig.~\ref{fig4}c); 
when $\delta_{\rm v}=-10\tau$ and $\delta_{\rm R}=-10\tau$, the robot moves towards the low-light-intensity regions near the edges of the arena where the speed is high and the rotational diffusion is low (Fig.~\ref{fig4}e).
This enhancement of the robot motion is further confirmed by the changes in the corresponding $\rho_0(r)$ (Figs.~\ref{fig4}f-h) and $d(r)$ (Figs.~\ref{fig4}i-k).

In Fig.~\ref{fig5}, we consider the case when the two sensorial delays have opposite signs, and compete with each other.
Figs.~\ref{fig5}a and \ref{fig5}b show $v(r)$ and $D_{\rm R}(r)$, respectively. 
Some sample trajectories are shown in Figs.~\ref{fig5}c and \ref{fig5}d: 
when $\delta_{\rm v}=+10\tau$ and $\delta_{\rm R}=-10\tau$, the robot is attracted towards the regions with low-light-intensity regions where the speed is high and the rotational diffusion is low (Fig.~\ref{fig5}c);
when $\delta_{\rm v}=-10\tau$ and $\delta_{\rm R}=+10\tau$, the robot is attracted towards the high-light-intensity regions where the speed is low and the rotational diffusion is high (Fig.~\ref{fig5}d).
These results are supported by $\rho_0(r)$ (Figs.~\ref{fig5}e-f) and $D(r)$ (Fig.~\ref{fig5}g-h).

\subsection{Differences between theory and simulations/experiments}\label{sec:tvs}

\begin{figure*}
\includegraphics[width=.75\textwidth]{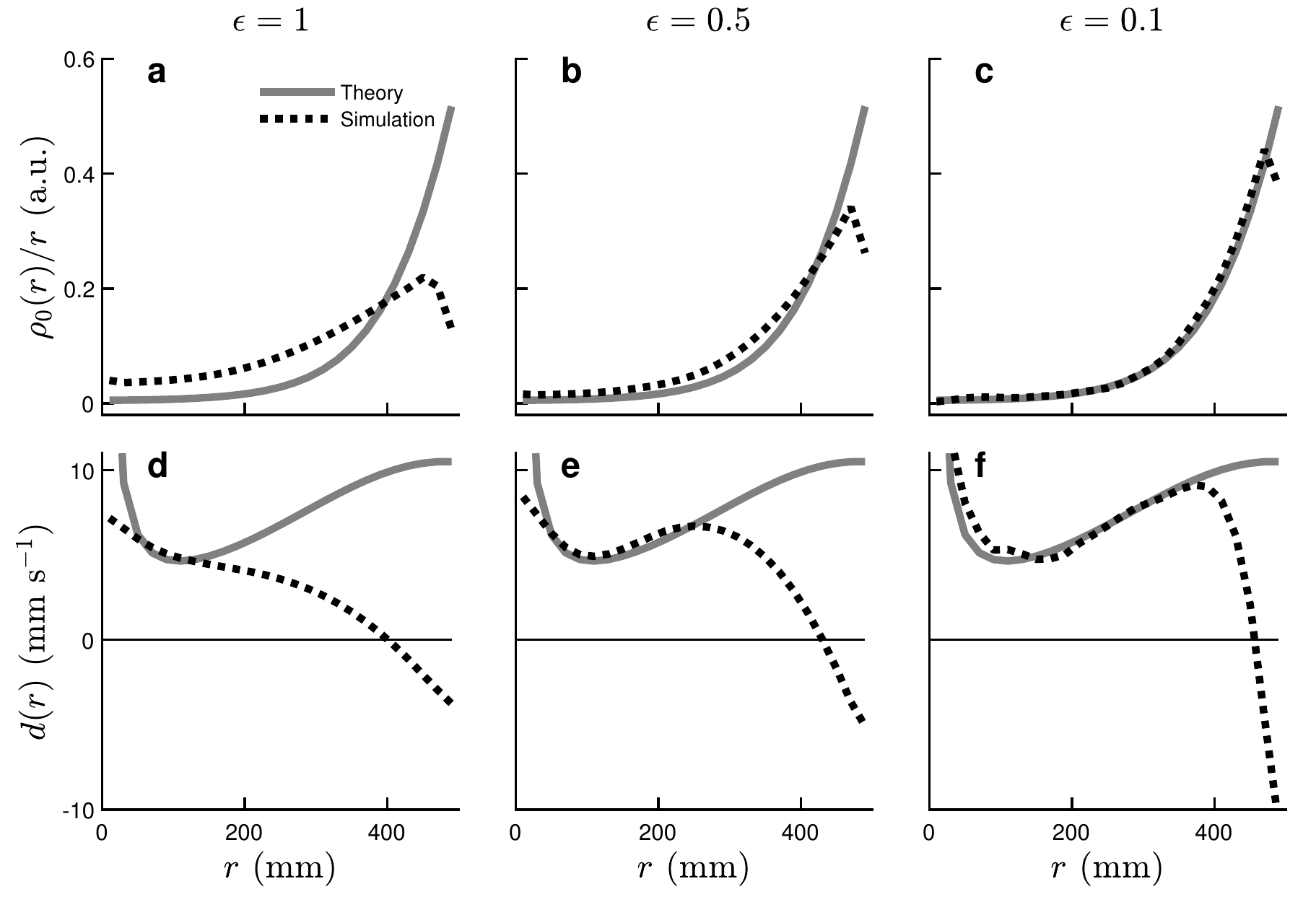}
\caption{
Convergence of the simulation results to the theoretical predictions for a simulated agent with sensorial delay in the speed.
The simulated (a-c) radial probability distribution $\rho_0(r)/r$ and (d-f) radial drift $d(r)$ converge towards the theoretical prediction as $\epsilon \to 0$ (while keeping the $\delta_{\rm v}/\tau$ and $\delta_{\rm R}/\tau$ constant).
}
\label{fig6}
\end{figure*}

In all data presented in Figs.~\ref{fig2}, \ref{fig3}, \ref{fig4}, and \ref{fig5}, we obtain a very good agreement between the experiments and simulations, while  there are certain discrepancies when it comes to the theory, particularly for the cases of the drift with negative delays. 
This can be explained taking into consideration that the theory assumes taking the value $\epsilon \to 0$ in SDDEs~(\ref{eq:SDDE}) (while keeping $\delta_{\rm v}/\tau$ and $\delta_{\rm R}/\tau$ constant), while the simulations and experiments are performed at the finite value of $\epsilon = 1$. 
We tested this hypothesis by running simulations where $\epsilon$ was taken towards zero while keeping the ratio between the time scales of the system the same as in the experiments. 
In Fig.~\ref{fig6}, the results can be observed for a simulated robot whose speed varies as a function of the intensity under the influence of a negative delay of $\delta_{\rm v}=-10\tau$. 
The radial probability distribution of the robot can be seen to converge towards the theoretical distribution when $\epsilon \to 0$ (Figs.~\ref{fig6}a-c). 
An even more significant change can be observed for the robot's radial drift:
while for $\epsilon = 1$ (Fig.~\ref{fig6}d) there is a significant difference between the simulated and theoretical radial drifts, this difference is significantly reduced when $\epsilon = 0.5$ (Fig.~\ref{fig6}e) and, even more, when $\epsilon = 0.1$ (Fig.~\ref{fig6}f).
Note also that in all cases there is a sharp drop in the simulated robot radial drift near the boundary, which is due to the fact that the theory does not account for the boundary's presence.

\section{Conclusions}

We have explored the role that sensorial delays play in determining the motion of an autonomous robot. Extending our previous work \cite{mijalkov2016engineering}, we have considered a phototactic robot whose speed and rotational diffusion depend on the local value of the light intensity. We have shown that the introduction of sensorial delays leads to an alteration of both the position probability distribution and the drift of the robot.
These results can be used to engineer the motion of autonomous agents using sensorial delays as well as to explain how multiple sensorial delays can interplay to obtain the desired behavior of a system.

\begin{acknowledgements}
We thank Lovisa Hagst\"{o}m, Erik Holmberg, Eliza Nord\'{e}n, Teodor Norrestad, Martin Selin and Lisa Sj\"{o}blom for performing an early version of the simulations presented in this work during their Bachelor Thesis, as well as Mite Mijalkov and Gilles Caprari for useful discussions.
This work was partially supported by the ERC Starting Grant ComplexSwimmers (Grant No. 677511).
JW's work was partially supported by the NSF grant DMS-1615045.
\end{acknowledgements}

\appendix

\section{Mathematical Derivation}\label{appendix}

Starting from SDDEs~(\ref{eq:SDDE}), we approximate them with a system of SDEs without delays by linearizing $x$ and $y$ as functions of time, and then $v$ and $\sigma$ as functions of $x$ and $y$.  
As a result, we obtain:
\begin{equation}\label{app2}
v\left(x_{t-c\epsilon^2}, y_{t-c\epsilon^2}\right) \approx v\left(x_t,y_t\right) - v_x\left(x_t,y_t\right)c\epsilon^2\dot{x}_t - v_y\left(x_t,y_t\right)c\epsilon^2\dot{y}_t 
\end{equation}
with $v_x$ and $v_y$ denoting the partial derivatives of $v$, and dots denoting time derivatives.  Substituting this expression into the first two equations of SDDEs~(\ref{eq:SDDE}), we obtain approximate versions of these equations:
\begin{equation}\label{app3}
    \begin{gathered}
    \begin{cases}
        \displaystyle \dot{x}_t = {1 \over \epsilon}v\cos\phi_t - c\epsilon v_x\cos\phi_t\dot{x}_t - c\epsilon v_y\cos\phi_t\dot{y}_t \\[10pt]
        \displaystyle \dot{y}_t = {1 \over \epsilon}v\sin\phi_t - c\epsilon v_x\sin\phi_t\dot{x}_t - c\epsilon v_y\sin\phi_t\dot{y}_t
        \end{cases}
    \end{gathered}
\end{equation}
From this point on, $v$, $v_x$ and $v_y$ are always evaluated at $(x_t,y_t)$ and we omit their arguments from the notation.
Eqs.~(\ref{app3}) constitute a system of linear equations for $\dot{x}_t$ and $\dot{y}_t$, whose solution is
\begin{equation}\label{app4}
    \begin{gathered}
    \begin{cases}
        \displaystyle \dot{x}_t = {1 \over \epsilon}v\cos\phi_t\left[1 + c\epsilon\left(v_x\cos\phi_t + v_y\sin\phi_t\right)\right]^{-1} \\[10pt]
        \displaystyle \dot{y}_t = {1 \over \epsilon}v\sin\phi_t\left[1 + c\epsilon\left(v_x\cos\phi_t + v_y\sin\phi_t\right)\right]^{-1}
        \end{cases}
    \end{gathered}
\end{equation}
For small $\epsilon$, we can approximate further, obtaining the first two equations of the system we will study:
\begin{equation}\label{app5}
    \begin{gathered}
    \begin{cases}
        \displaystyle \dot{x}_t = {1 \over \epsilon}v\cos\phi_t\left[1 - c\epsilon\left(v_x\cos\phi_t + v_y\sin\phi_t\right)\right] \\[10pt]
        \displaystyle \dot{y}_t = {1 \over \epsilon}v\sin\phi_t\left[1 - c\epsilon\left(v_x\cos\phi_t + v_y\sin\phi_t\right)\right]
        \end{cases}
    \end{gathered}
\end{equation}
To obtain the third equation, we start from a similar approximation of the function $\sigma$:
\begin{equation}\label{app6}
\sigma\left(x_{t - k\epsilon^2}, y_{t-k\epsilon^2}\right) \approx \sigma\left(x_t,y_t\right) - \sigma_x\left(x_t,y_t\right)k\epsilon^2\dot{x}_t - \sigma_y\left(x_t,y_t\right)k\epsilon^2\dot{y}_t .
\end{equation}
We further approximate the expression on the right-hand side, replacing $\dot{x}_t$ and $\dot{y}_t$ by their leading order terms\footnote{Including higher order terms in the approximations for $\dot{x}_t$ and $\dot{y}_t$, substituted into Eqs.~(\ref{app6}), would give rise to terms of order $\epsilon$ in the equation for $\phi_t$ (Eq.~(\ref{app8})).  As can be seen from the asymptotic analysis that follows, this would not change the equation obtained in the $\epsilon \to 0$ limit. }
from Eqs.~(\ref{app5}):
\begin{equation}\label{app5}
    \begin{gathered}
    \begin{cases}
        \displaystyle \dot{x}_t \approx {1 \over \epsilon}v\cos\phi_t \\[10pt]
        \displaystyle \dot{y}_t \approx {1 \over \epsilon}v\sin\phi_t
        \end{cases}
    \end{gathered}
\end{equation}
to obtain the third equation of the approximate system:
\begin{equation}\label{app8}
d\phi_t =\left[ {1 \over \epsilon}\sigma\left(x_t,y_t\right) - k\sigma_xv\cos\phi_t - k\sigma_y v\sin\phi_t \right]\,dW_t .
\end{equation}

In order to study the limiting behavior of the process $(x_t,y_t)$, we introduce the associated (backward) Kolmogorov operator.\footnote{The general rule is the following: consider 
a system of SDE
$$
dx^i_t = b^i(x_t)\,dt + \sum_{\alpha=1}^l\sigma^i_{\alpha}(x_t)\,dW_t^{\alpha}
$$
where $i = 1, \dots k$ and $W^1, \dots W^l$ are independent Wiener processes.  The generator is then the differential operator
$$
L = {1 \over 2}\sum_{i,j = 1}^ka^{ij}(x){\partial^2 \over \partial x^i\partial x^j} + \sum_{i=1}^kb^i(x){\partial \over \partial x^i}
$$
where the $a^{ij}$ are matrix elements of the matrix $a = \sigma\sigma^T$, i.e. $a^{ij} = \sum_{\alpha}\sigma^i_{\alpha}\sigma^j_{\alpha}$.  
Consult \cite{oksendal2003stochastic} for more details.}  
In our case,
\begin{align}\label{app9}
L = 
&{1 \over 2}\left[{1 \over \epsilon}\sigma - k\sigma_xv\cos \phi - k\sigma_yv\sin\phi\right]^2\partial^2_{\phi\phi} \cr
&+ \left({1 \over \epsilon}v\cos \phi - cvv_x\cos^2\phi -cvv_y\cos\phi\sin\phi\right)\partial_x  \cr
&+ \left({1 \over \epsilon}v\sin\phi - cvv_x\sin\phi\cos\phi - cvv_y\sin^2\phi\right)\partial_y. \cr
\end{align}
Considering the corresponding backward Kolmogorov equation for a function $p(t, x, y, \phi)$,
\begin{equation}\label{app10}
\partial_t \rho = L\rho,
\end{equation}
we have
\begin{equation}\label{app11}
L = \epsilon^{-2}L_{-2} + \epsilon^{-1}L_{-1} + L_0,
\end{equation}
where
\begin{align}\label{app12}
L_{-2} = 
& {1 \over 2}\sigma^2\partial^2_{\phi\phi} \cr
L_{-1} = 
& \left(-k\sigma\sigma_x v\cos\phi - k\sigma\sigma_y v\sin\phi\right)\partial^2_{\phi\phi} \cr
& + v\cos\phi\partial_x + v\sin\phi\partial_y \cr
L_0 = 
& \left({1 \over 2}k^2\sigma_x^2v^2\cos^2\phi + k^2\sigma_x\sigma_yv^2\cos\phi\sin\phi \right.\cr
& \left. + {1 \over 2}k^2\sigma_y^2v^2\sin^2\phi\right)\partial^2_{\phi\phi} \cr
& - \left(cvv_x\cos^2\phi + cvv_y\cos\phi\sin\phi\right)\partial_x \cr
& - \left(cvv_x\sin\phi\cos\phi + cvv_y\sin^2\phi\right)\partial_y \cr
\end{align}
We write the function $\rho$ as a formal series in powers of $\epsilon$:
\begin{equation}\label{app13}
\rho = \rho_0 + \epsilon \rho_1 + \epsilon^2 \rho_2 + \dots,
\end{equation}
substitute it into Eq.~(\ref{app10}), and equate coefficients of the same powers of $\epsilon$ on both sides of the resulting equation.  The goal is to obtain a differential equation for $\rho_0$, which (in view of Eq.~(\ref{app13})) is the limit of $\rho$ as $\epsilon \to 0$.

In order $\epsilon^{-2}$, we obtain
\begin{equation}\label{app14}
L_{-2}\rho_0 = 0,
\end{equation}
which implies $\partial^2_{\phi\phi}\rho_0 = 0$.  While the general solution of this equation is an affine function of $\phi$, that is, has a form $a(x,y) + b(x,y)\phi$, we choose a solution $\rho_0(x,y)$ which does not depend on $\phi$, since we expect that the limiting equation does not involve the fast variable $\phi$.  

In order $\epsilon^{-1}$, we have
\begin{equation}\label{app15}
L_{-2}\rho_1 = -L_{-1}\rho_0,
\end{equation}
which implies the equation
\begin{equation}\label{app16}
\partial^2_{\phi\phi}\rho_1 = -{2v \over \sigma^2}\partial_x\rho_0\cos\phi - {2v \over \sigma^2}\partial_y\rho_0\sin\phi
\end{equation}
whose solution, periodic in $\phi$, is 
\begin{equation}\label{app17}
\rho_1 = {2v \over \sigma^2}\partial_x\rho_0\cos\phi + {2v \over \sigma^2}\partial_y\rho_0\sin\phi.
\end{equation}

In order $\epsilon^0$, we obtain
\begin{equation}\label{app18}
\partial_t\rho_0 = L_{-2}\rho_2 + L_{-1}\rho_1 + L_0\rho_0.
\end{equation}
This can be rewritten as  $\partial_t\rho_0 - L_{-1}\rho_1 - L_0\rho_0 = L_{-2}\rho_2$, which implies that the function 
$\partial_t\rho_0 - L_{-1}\rho_1 - L_0\rho_0$ belongs to the range of the operator $L_{-2}$, and is thus orthogonal to the null space of the adjoint operator $L_{-2}^*$.\footnote{This is a general fact about linear operators on Hilbert spaces.  A discussion in the present context can be found e.g. in \cite{pavliotis2008multiscale}.}
$L_{-2}$ is considered here as an operator in the variable $\phi$.  Since $\sigma$ does not depend on $\phi$, $L_{-2}^* = L_{-2}$ and the null space of this operator is spanned by the constant function $1$.  The orthogonality relation becomes
\begin{equation}\label{app19}
\partial_t \rho_0 = {1 \over 2\pi}\int_{-\pi}^{\pi}\left(L_0\rho_0 +L_{-1}\rho_1\right)\,d\phi.
\end{equation}
Substituting
\begin{equation}\label{app20}
L_0\rho_0 = -\left(cvv_x\cos^2\phi\partial_x\rho_0 +cvv_y\sin^2\phi\partial_y\rho_0\right)
\end{equation}
and 
\begin{align}\label{app21}
L_{-1}\rho_1 = 
& {2kv^2 \over \sigma}\left(\sigma_x\cos\phi + \sigma_y\sin\phi\right)\left(\partial_x\rho_0\cos\phi + \partial_y\rho_0\sin\phi\right) \cr
&+v\cos\phi\,\partial_x\left( {2v \over \sigma^2}\partial_x\rho_0\cos\phi + {2v \over \sigma^2}\partial_y\rho_0\sin\phi \right) \cr
&+ v\sin\phi\,\partial_y\left( {2v \over \sigma^2}\partial_x\rho_0\cos\phi + {2v \over \sigma^2}\partial_y\rho_0\sin\phi \right), \cr
\end{align}
and using the trigonometric integrals
\begin{equation}\label{app22}
{1 \over 2\pi}\int_{-\pi}^{\pi}\cos^2\phi\,d\phi = {1 \over 2\pi}\int_{-\pi}^{\pi}\sin^2\phi\,d\phi = {1 \over 2}
\end{equation}
and
\begin{equation}\label{app23}
{1 \over 2\pi}\int_{-\pi}^{\pi}\cos\phi\sin\phi\,d\phi = 0,
\end{equation}
we obtain
\begin{align}\label{app24}
\partial_tp_0 = L\rho_0 =  
& -{1 \over 2}cvv_x\partial_x\rho_0 - {1 \over 2}cvv_y\partial_y\rho_0 \cr
& + k{v^2 \over \sigma}\sigma_x\partial_x\rho_0 +  k{v^2 \over \sigma}\sigma_y\partial_y\rho_0 \cr
& + v\partial_x\left({v \over \sigma^2}\partial_x\rho_0\right) + v\partial_y\left({v \over \sigma^2}\partial_y\rho_0\right) \cr
\end{align}
or, in vector notation,
\begin{equation}\label{app24}
\partial_t\rho_0 = -{1 \over 2}cv\nabla v\cdot \nabla \rho_0 + k{v^2 \over \sigma}\nabla\sigma\cdot\nabla \rho_0 + v\nabla\cdot\left({v \over \sigma^2}\nabla \rho_0\right).
\end{equation}
This is the limiting backward Kolmogorov equation for a function $\rho_0$ of the variables $x$ and $y$, from which we obtain the system of SDEs, satisfied by the processes $x_t$ and $y_t$:
\begin{equation}\label{app26}
    \begin{gathered}
    \begin{cases}
        \displaystyle dx_t = \left[-{1 \over 2}cvv_x + k{v^2 \over \sigma}\sigma_x + v\partial_x\left({v \over \sigma^2}\right)\right]\,dt + \sqrt{2}{v \over \sigma}\,dW_t^{(1)} \\[10pt]
        \displaystyle dy_t = \left[-{1 \over 2}cvv_y + k{v^2 \over \sigma}\sigma_y + v\partial_y\left({v \over \sigma^2}\right)\right]\,dt + \sqrt{2}{v \over \sigma}\,dW_t^{(2)} \cr
        \end{cases}
    \end{gathered}
\end{equation}
where $W^{(1)}$ and $W^{(2)}$ are independent Wiener processes.\footnote{This step is a reversal of the previous operation by which we obtained a Kolmogorov equation from an SDE system. Having passed to the limit at the level of Kolmogorov equations, we revert back to the corresponding SDEs.}  Note that in the case when $\sigma$ is identically equal to $1$, we obtain the system studied previously in Ref.~\cite{mijalkov2016engineering}.  

Passing to formal adjoints, we obtain the associated forward Kolmogorov (i.e., Fokker-Planck) equation:
\begin{align}\label{app27}
\partial_t\rho_0 = L^*\rho_0 =  
& {1 \over 2}c\partial_x\left(vv_x\rho_0\right) +  {1 \over 2}c\partial_y\left(vv_y\rho_0\right) \cr
& - k\partial_x\left({v^2 \over \sigma}\sigma_x \rho_0\right) - k\partial_y\left({v^2 \over \sigma}\sigma_y \rho_0\right) \cr
&- \partial_x\left(v\partial_x\left({v \over \sigma^2}\right)\rho_0\right) - \partial_y\left(v\partial_y\left({v \over \sigma^2}\right)\rho_0\right) \cr
& + \partial^2_{xx}\left({v^2 \over \sigma^2}\rho_0\right) + \partial^2_{yy}\left({v^2 \over \sigma^2}\rho_0\right), \cr
\end{align}
which, in vector form, becomes
\begin{equation}\label{app28}
\partial_t \rho_0 = {1 \over 2}c\nabla\cdot\left(\rho_0 v\nabla v\right) - k\nabla\cdot\left(\rho_0{v^2 \over \sigma}\nabla\sigma\right) + \nabla\cdot\left({v \over \sigma^2}\nabla\left(v\rho_0\right)\right).
\end{equation}
If the system possesses a stationary probability density $\rho_0$, then $\rho_0$ has to satisfy the stationary Fokker-Planck equation
\begin{equation}\label{app29}
L^*\rho_0 = 0~.
\end{equation}

We are now going to consider some special cases, where the solutions of the stationary Fokker-Planck equation actually satisfy a stronger condition. 
Ref.~\cite{birrel2017homogenization} explains in detail that in these situations we actually obtain an equilibrium distribution, i.e., a stationary distribution satisfying the detailed balance condition.

\subsubsection{Constant rotational diffusion}

Suppose $\sigma$ is constant, which correspond to the special case we studied in Ref.~\cite{mijalkov2016engineering}.
In this case the stationary Fokker-Planck equation becomes
\begin{equation}\label{app30}
{1 \over 2}c\nabla\cdot\left(\rho_0 v\nabla v\right) + {1 \over \sigma^2}\nabla\cdot\left(v\nabla\left(v\rho_0\right)\right) = 0.
\end{equation}
We search for a solution of 
\begin{equation}\label{app31}
{1 \over 2}c\rho_0 v\nabla v + {1 \over \sigma^2}v \nabla\left(v\rho_0\right) = 0,
\end{equation}
which can be rewritten as
\begin{equation}\label{app32}
{\nabla\left(v\rho_0\right) \over v\rho_0} = -{\sigma^2 \over 2}c{\nabla v\over v}
\end{equation}
and integrated to yield
\begin{equation}\label{app33}
\rho_0 = Bv^{-\left(1 + {\sigma^2c \over 2}\right)}.
\end{equation}
In a bounded domain, a positive value of $B$ can always be chosen, so as to make $\rho_0$ a probability distribution.  For $c > -{2 \over \sigma^2}$, the points $(x,y)$ with smaller values of $v$ are preferred by this distribution; for $c < -{2 \over \sigma^2}$ the tendency is reversed.

\subsubsection{Constant speed}

Suppose $v$ is constant.  In this case the stationary Fokker-Planck equation becomes
\begin{equation}\label{app34}
-k\nabla\cdot\left(\rho_0{v^2 \over \sigma}\nabla\sigma\right) + \nabla\cdot\left({v \over \sigma^2}\right)\nabla\left(v\rho_0\right)= 0.
\end{equation}
A function $\rho_0$ will satisfies this equation if it satisfies the equation
\begin{equation}\label{app35}
-kv^2{\rho_0 \over \sigma}\nabla\sigma + {v \over \sigma^2}\nabla\left(v\rho_0\right) = 0.
\end{equation}
For this, it is enough to find a solution of 
\begin{equation}\label{app36}
{\nabla\left(v\rho_0\right) \over v\rho_0} = k\sigma\nabla\sigma,
\end{equation}
i.e.,
\begin{equation}\label{app37}
\nabla\log\left(v\rho_0\right) = {1 \over 2}k\nabla\left(\sigma^2\right),
\end{equation}
which has a general solution 
\begin{equation}\label{app38}
\rho_0 = B\exp\left({1 \over 2}k\sigma^2\right).
\end{equation}
This can be normalized to become a probability density if the point $(x,y)$ is restricted to a bounded domain.  For $k > 0$, it shows that the particle is more likely to be found in the region where its rate of rotation is bigger.  For $k < 0$ it has the opposite tendency.

\subsubsection{Speed proportional to the rotational diffusion}

Yet another case in which the stationary Fokker-Planck equation can be integrated explicitly is the case when $v / \sigma^2$ is a constant, i.e., when the speed is proportional to the rotational diffusion. The calculation is straightforward.

\subsubsection{Radial coordinates}

Suppose both $v$ and $\sigma$ are functions of $r = \sqrt{x^2 + y^2}$, which is the special case of the experiment we performed. We are going to find an SDE satisfied by $r_t = \sqrt{x_t^2 + y_t^2}$ and use it to derive the stationary distribution of the particle's distance from the origin.  To this end, we use the It\^o formula for the function  $r = \sqrt{x^2 + y^2}$,
\begin{equation}\label{app39}
dr_t = {x_t \over r_t}\,dx_t + {y_t \over r_t}\,dy_t + {1 \over 2}{r_t^2 - x_t^2 \over r_t^3}\left(dx_t\right)^2 + {1 \over 2}{r_t^2 - y_t^2 \over r_t^3}\left(dy_t\right))^2,
\end{equation}
and substitute the expressions for $dx_t$ and $dy_t$ from Eqs.~(\ref{app26}).  Noting that for a function $f(x,y)$ we have
$f_r = f_xx_r + f_yy_r = {x \over r}f_x + {y \over r}f_y$, and using the fact that ${x_t \over r_t}\,dW_t^{(1)} + {y_t \over r_t}\,dW_t^{(2)}$ is a differential of a Wiener process, which we will denote by $\tilde{W}_t$, we obtain
\begin{equation}\label{app40}
dr_t = \left(-{1 \over 2}cvv_r + k{v^2\sigma_r \over \sigma} + v\left({v \over \sigma^2}\right)_r + {1 \over r}{v^2 \over \sigma^2}\right)\,dt + \sqrt{2}{v \over \sigma}\,d\tilde{W}_t.
\end{equation}
Denoting the drift and the noise coefficients in the above SDE by $b(r)$ and $s(r)$ respectively, we have the standard formula for the density $\rho_0$ of the stationary distribution of $r$:
\begin{equation}\label{app41}
\rho_0(r) = {B\over s^2(r)}\exp\left(\int_{r_0}^r{2b(u) \over s(u)^2}\right)\,du.
\end{equation}
Here, $r_0 > 0$ is the minimal distance of the particle from the origin, allowed by the experimental restrictions \cite{birrel2017homogenization} and $B$ is the constant, normalizing the integral of $g$ to $1$.  The integrand ${2b \over s^2}$ can be written explicitly as
\begin{equation}\label{app42}
{2b \over s^2} = -{1 \over 2}c{\sigma^2v_r \over v} + k\sigma\sigma_r + {\sigma^2 \over v}\left({v \over \sigma^2}\right)_r + {1 \over r}.
\end{equation}

Here is a short derivation of Eq.~(\ref{app41}):  according to the general rule, the generator of the process $r_t$ is given by
\begin{equation}\label{app43}
L = {1 \over 2}s(r)^2 {d^2 \over dr^2} + b(r){d \over dr};
\end{equation}
the adjoint generator is thus acting on functions of $r$ according to the formula
\begin{equation}\label{app44}
(L^*f)(r) = {1 \over 2}{d^2 \over dr^2}\left(s(r)^2f(r)\right) - {d \over dr}\left(b(r)f(r)\right);
\end{equation}
to find the stationary density we solve the equation $L^*\rho_0 = 0$, searching for a solution, satisfying the (stronger) equation
\begin{equation}\label{app45}
{1 \over 2}{d \over dr}\left(s(r)^2\rho_0(r)\right) - \left(b(r)\rho_0(r)\right) = 0;
\end{equation}
to solve this first-order ODE, we substitute $g = s^2\rho_0$ and obtain
\begin{equation}\label{app46}
{1 \over 2}{dg \over dr} - {b \over s^2}g = 0,
\end{equation}
which can be solved by separation of variables, i.e.,
\begin{equation}\label{app47}
g = B\exp\left(\int {2b(r) \over s^2(r)}\, dr\right)
\end{equation}
where $B$ is a constant; from this formula, we obtain 
\begin{equation}\label{app48}
\rho_0(r) = {B \over s^2(r)}\exp\left(\int {2b(r) \over s^2(r)}\, dr\right)
\end{equation}
where $B$ has to be adjusted to make the integral of $\rho_0$ equal $1$.


\end{document}